\documentclass[aps,prl,10pt,superscriptaddress,twocolumn,floatfix,showkeys]{revtex4-2}
\usepackage{amsmath, amssymb, amsfonts, multirow}
\usepackage[USenglish]{babel}
\usepackage[T1]{fontenc}
\usepackage{graphicx}
\usepackage{natbib}
\usepackage{textcomp}
\usepackage[dvipsnames]{xcolor}
\usepackage[utf8]{inputenc}
\usepackage[version=4]{mhchem}
\usepackage{makecell}
\usepackage{mathptmx}
\usepackage{etoolbox}
\usepackage{xcolor}
\usepackage{soul}
\usepackage{float}
\usepackage{calrsfs}
\usepackage{siunitx}
\let\svqty\qty
\usepackage{physics}
\let\qty\svqty
\usepackage{lineno}
\usepackage{silence}
\usepackage{hyperref}
\usepackage{orcidlink}

\definecolor{darkgoldenrod}{rgb}{0.72, 0.53, 0.04}

\AtBeginDocument{\RenewCommandCopy\qty\SI}
\ExplSyntaxOn
\msg_redirect_name:nnn { siunitx } { physics-pkg } { none }
\ExplSyntaxOff

\newcommand{\muB}{\mu_{\text{B}}}

\renewcommand{\appendixname}{Supplementary Information}
\begin{document}
\title{Telecom-Wavelength-Compatible Quantum Information Transcription Using Nitrogen-Vacancy Centers}

\author{\orcidlink{0009-0007-4399-7962}{B.~G\"obly\"os}}
\thanks{These authors contributed equally to this work.}
\affiliation{{Department of Physics, Institute of Physics, Budapest University of Technology and Economics, M\H{u}egyetem rkp. 3., H-1111 Budapest, Hungary}}
\affiliation{{HUN-REN-BME Condensed Matter Research Group, Budapest University of Technology and Economics, M\H{u}egyetem rkp. 3., H-1111 Budapest, Hungary}}

\author{\orcidlink{0000-0002-0826-7928}{S.~Kollarics}}
\thanks{These authors contributed equally to this work.}
\affiliation{{Institute for Solid State Physics and Optics, HUN-REN Wigner Research Centre for Physics, PO. Box 49, H-1525, Hungary}}
\affiliation{{Department of Physics, Institute of Physics, Budapest University of Technology and Economics, M\H{u}egyetem rkp. 3., H-1111 Budapest, Hungary}}

\author{\orcidlink{0009-0003-8889-6509}{R.~Kucsera}}
\affiliation{{Department of Physics, Institute of Physics, Budapest University of Technology and Economics, M\H{u}egyetem rkp. 3., H-1111 Budapest, Hungary}}
\affiliation{{Qutility @ Faulhorn Labs, Budafoki \'{u}t 91-93, H-1117 Budapest, Hungary}}

\author{\orcidlink{0009-0005-7364-9518}{D.~Plitt}}
\affiliation{{Department of Physics, Institute of Physics, Budapest University of Technology and Economics, M\H{u}egyetem rkp. 3., H-1111 Budapest, Hungary}}
\affiliation{Faculty of Physics and CENIDE, University of Duisburg-Essen, Lotharstraße 1, 47057 Duisburg, Germany}

\author{K.~Koltai}
\affiliation{{Department of Physics, Institute of Physics, Budapest University of Technology and Economics, M\H{u}egyetem rkp. 3., H-1111 Budapest, Hungary}}

\author{\orcidlink{0000-0002-3513-328X}{L.~Forr\'{o}}}
\affiliation{Stavropoulos Center for Complex Quantum Matter, Department of Physics and Astronomy, University of Notre Dame, Notre Dame, Indiana 46556, USA}

\author{\orcidlink{0000-0003-1472-0482}{B.~G.~M\'{a}rkus}}
\email[Corresponding author: ]{bmarkus@nd.edu}
\affiliation{Stavropoulos Center for Complex Quantum Matter, Department of Physics and Astronomy, University of Notre Dame, Notre Dame, Indiana 46556, USA}

\author{\orcidlink{0000-0001-9822-4309}{F.~Simon}}
\email[Corresponding author: ]{simon.ferenc@ttk.bme.hu}
\affiliation{{Department of Physics, Institute of Physics, Budapest University of Technology and Economics, M\H{u}egyetem rkp. 3., H-1111 Budapest, Hungary}}
\affiliation{{HUN-REN-BME Condensed Matter Research Group, Budapest University of Technology and Economics, M\H{u}egyetem rkp. 3., H-1111 Budapest, Hungary}}
\affiliation{{Institute for Solid State Physics and Optics, HUN-REN Wigner Research Centre for Physics, PO. Box 49, H-1525, Hungary}}
\affiliation{Stavropoulos Center for Complex Quantum Matter, Department of Physics and Astronomy, University of Notre Dame, Notre Dame, Indiana 46556, USA}


\begin{abstract}
    Nitrogen-vacancy (NV) centers in diamond are a leading platform for solid-state quantum sensing and quantum information processing. While most optical studies rely on the visible fluorescence associated with the triplet transitions, the infrared singlet transition near $1042$ nm, which is typically considered dark within the singlet manifold of the NV optical cycle, provides an alternative optical channel. Here, we report wavelength-resolved optically detected magnetic resonance (ODMR) measurements of this infrared emission. We directly observe ODMR contrast in the $1042$ nm emission and analyze its dependence on the magnetic field. The field-dependent spectral dispersion of the ODMR signal demonstrates that the spin-state information encoded in the NV center is transcribed to the infrared singlet emission through the spin-selective intersystem crossing, in close analogy to the visible fluorescence readout. These results establish infrared ODMR as a high-fidelity optical readout pathway. Crucially, by extending spin-state transcription directly into the $1300-1600$ nm range, this work demonstrates a direct, conversion-free interface between diamond spin-qubits and standard telecommunication infrastructure, bypassing the efficiency bottlenecks of active frequency conversion and benefiting from the already well-developed technologies in this range of the electromagnetic spectrum.
\end{abstract}

\maketitle
\date{}

\section{Introduction}

Nitrogen-vacancy (NV) centers in diamond constitute one of the most prominent solid-state platforms for quantum sensing, quantum communication, and quantum information processing, owing to their optically addressable spin states and long spin coherence times even at room temperature \cite{Doherty2013, Schirhagl2014, casola_probing_2018, kollarics_terahertz_2024}. In the negatively charged NV$^{-}$ center, the electronic ground state is a spin triplet whose spin sublevels can be initialized and read out optically through spin-dependent fluorescence in the visible spectral range. Optical excitation populates the excited triplet state, while a spin-selective intersystem crossing (ISC) to intermediate singlet states provides the mechanism that converts spin information into an intensity contrast in the emitted fluorescence \cite{Gruber1997, Jelezko2006}. This spin-dependent optical cycle forms the basis of optically detected magnetic resonance (ODMR), which has enabled highly sensitive magnetometry, thermometry, and electric-field sensing using individual NV centers and ensembles \cite{Taylor2008, Rondin2014, du_control_2017, Arunkumar2021PRX}.

Most optical detection schemes rely on the broadband visible fluorescence between approximately $637-800$~nm associated with the triplet transition. However, the NV optical cycle also contains a singlet manifold that gives rise to an infrared transition near $1042$~nm connecting the ${}^1E$ and ${}^1A_1$ singlet states \cite{Rogers2008, Acosta2010}. This transition is typically regarded as a \emph{dark} channel of the optical cycle because it does not contribute to the conventional fluorescence readout used in most ODMR experiments. Nevertheless, the population of the singlet manifold is governed by the same spin-dependent ISC processes that enable visible ODMR contrast, implying that spin information should also be encoded in the infrared emission.

The infrared optical channel of the negatively charged nitrogen-vacancy (NV$^{-}$) center originates from transitions within the singlet manifold of the NV optical cycle. The existence of this infrared emission was first established by Rogers \textit{et al}., who identified a zero-phonon line (ZPL) near $1046$~nm corresponding to the $^{1}E \to {}^{1}A_{1}$ singlet transition and studied its response to magnetic field and uniaxial stress \cite{Rogers2008}. Subsequent work clarified the role of this transition in the spin-dependent ISC that enables optical spin polarization and readout. In particular, Acosta \textit{et al}. investigated the optical properties and lifetimes of the singlet levels and demonstrated that the population of the singlet manifold is strongly spin dependent, making the infrared transition sensitive to the NV spin state \cite{Acosta2010}. Building on this understanding, Jensen \textit{et al}. demonstrated magnetometry using infrared absorption on the singlet transition at $1042$~nm, where optically detected magnetic resonance was observed via changes in infrared transmission through the diamond \cite{Jensen2014}. These studies established that spin information encoded in the NV center can be accessed through the singlet infrared transition, providing an alternative optical readout channel complementary to conventional visible fluorescence detection. 

\begin{figure*}[htp]
        \centering
        \includegraphics[width=\linewidth]{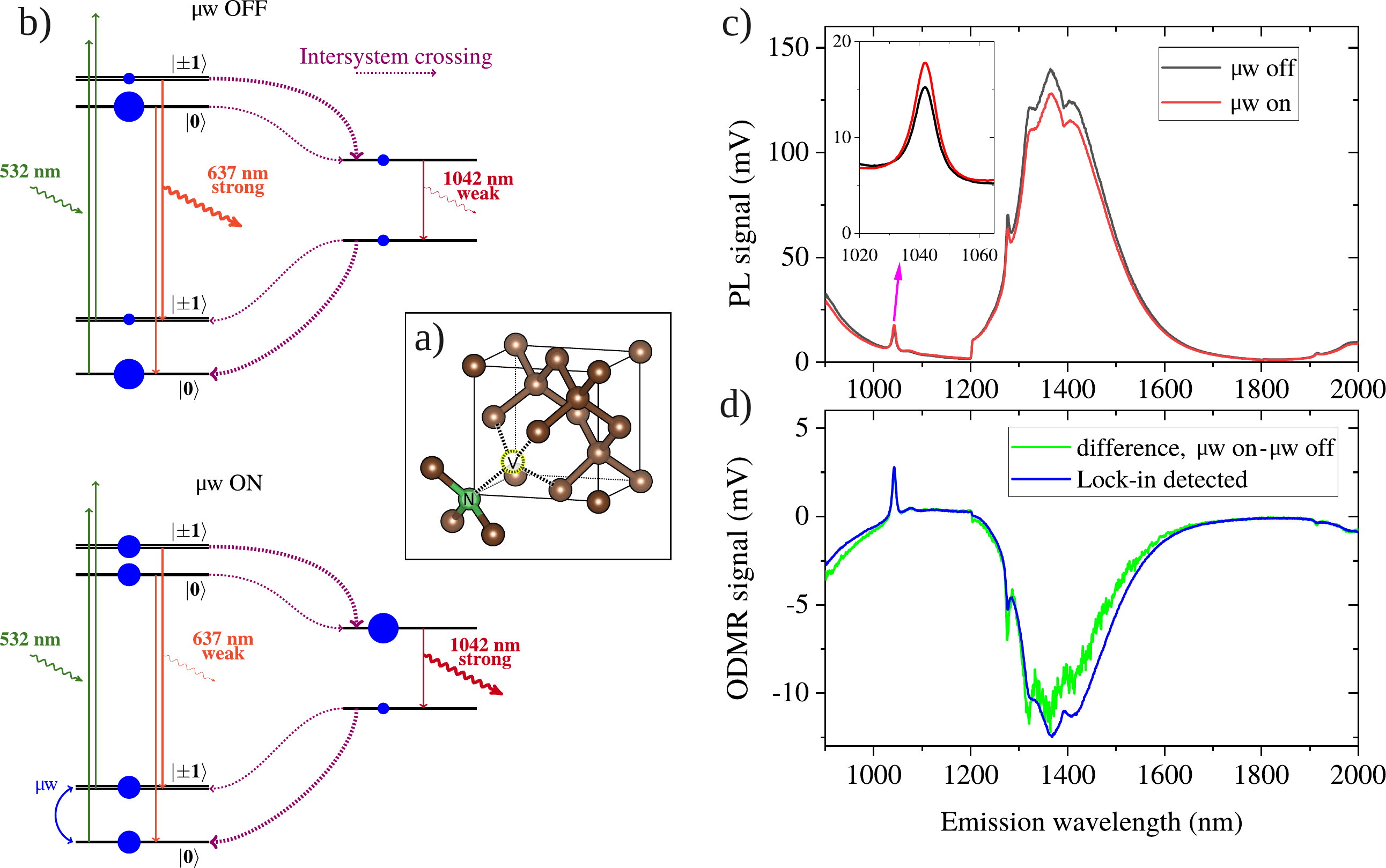}
        \caption{a) Structure of the diamond NV center, b) its Jablonski diagram, which shows the intermediate near-infrared transition between the singlet levels, c) the observed NIR emission using a $600$ nm long-pass filter with the microwave on/off, and d) the difference of the two spectra and the ODMR signal detected using the lock-in technique. The second-order diffraction of the visible spectrum appears in the near infrared detection with the characteristic ZPL of ($2\times 638~\text{nm}=$) $1276~\text{nm}$. The near infrared emission of the singlet states with the characteristic ZPL at $1042~\text{nm}$ increases while the visible emission decreases due to the microwave irradiation.}
        \label{fig:Fig1_structure_Jablonski_data}
\end{figure*}

Interest in infrared optical channels of NV centers has recently increased due to their potential relevance for hybrid quantum networks and long-distance quantum communication and encryption. While visible NV fluorescence ($637$ nm) requires complex, lossy quantum frequency down-conversion (QFC) to reach optimal telecom wavelengths for long-haul transmission \cite{Albrecht2013, Bernien2013}, the singlet infrared channel offers a hardware-agnostic alternative. We show that the NV center naturally encodes its quantum state into a broad infrared continuum. This enables a direct interface with existing fiber networks, potentially simplifying the architecture of future hybrid quantum repeaters by eliminating the need for intermediary conversion stages. Therefore, understanding whether and how spin information is preserved in the singlet infrared emission is an important step toward alternative readout schemes and toward interfaces connecting NV-based quantum sensors or nodes with telecom photonic technologies. Much as the IR emission from the singlet transition has been established, to our knowledge, the ODMR activity has not been studied in depth with wavelength resolution, especially not up to the telecom-wavelength relevant ranges, including the $1100-1500$ nm range.

We reveal a near-infrared window into the spin physics of nitrogen-vacancy (NV) centers in diamond by performing wavelength-resolved optically detected magnetic resonance (ODMR) on their infrared singlet emission. Focusing on the $1042$ nm zero-phonon line, we directly observe ODMR contrast and track its evolution with the magnetic field. The wavelength-dependent ODMR response shows that spin information from the NV center is faithfully transferred into the infrared singlet transition through spin-selective ISC. These findings establish infrared ODMR as a powerful alternative readout channel and point toward fiber-compatible, near-infrared NV technologies for future quantum sensing and information applications.

\section{Results and Discussion}

Fig.~\ref{fig:Fig1_structure_Jablonski_data}a shows the electronic structure of the diamond NV center together with its Jablonski diagram in Fig.~\ref{fig:Fig1_structure_Jablonski_data}b, which displays the photophysical processes relevant for optically detected magnetic resonance (ODMR). The negatively charged nitrogen-vacancy (NV$^-$) center consists of a substitutional nitrogen atom adjacent to a vacancy in the diamond lattice, forming a spin-triplet ground state ($^3A_2$) and an optically excited spin-triplet state ($^3E$), coupled via spin-selective ISC to intermediate singlet states ($^1A_1$~and~$^1E$) \cite{Doherty2013, Schirhagl2014}.

ODMR is based on the change of fluorescence intensity of NV centers under resonant microwave excitation. Under continuous optical pumping (typically $532$~nm), the system is polarized into the $m_{{S}}=0$ ground-state sublevel due to spin-dependent ISC, resulting in a steady fluorescence signal.

The ODMR contrast arises from the higher ISC probability of the $m_{{S}}=\pm1$ states compared to $m_{{S}}=0$. Microwave excitation transfers population into the $m_{{S}}=\pm1$ states, which preferentially decay through the singlet manifold, leading to a reduction of the visible fluorescence intensity \cite{Gruber1997, Jelezko2006}.

Although the singlet manifold is usually considered \emph{dark} in the visible range, it gives rise to a weak near-infrared transition ($^1E \to ^1A_1$) with a zero-phonon line at ${\sim}1042$~nm. This emission becomes observable when population accumulates in the singlet states, for example, under ODMR conditions \cite{Rogers2008, Acosta2010}.

Since the spin-dependent ISC determines the singlet population, microwave-driven spin redistribution modulates both the visible fluorescence and the NIR emission. Therefore, the infrared signal also carries ODMR contrast and provides an alternative optical readout channel \cite{Robledo2011}. Consequently, the detection of ODMR via the singlet-singlet transition provides additional insight into the internal population dynamics of the NV center, allowing for probing processes that are otherwise hidden in conventional fluorescence measurements. In Fig.~\ref{fig:Fig1_structure_Jablonski_data}b, we show the photoluminescence spectrum of the NV center in the $900-2000$~nm range, measured with and without microwave irradiation. A $600$~nm long-pass filter was placed before the monochromator to block the excitation light and the shorter-wavelength fluorescence. We used $2.874$ GHz for the microwave irradiation frequency that matches the $m_S=0\rightarrow m_S=\pm 1$ transition. 

The main feature is the emission around $1042$~nm, which corresponds to the zero-phonon line of the singlet--singlet transition. In addition, strong peaks appear above $1200$~nm. These are not real NIR features but originate from higher diffraction orders of the grating. They are caused by the strong visible fluorescence of the NV centers in the $600-1000$~nm range, which appears at longer wavelengths in higher orders. Above about $1800$~nm, even the third diffraction order becomes visible. Since the visible emission is much stronger than the NIR one, second-order diffraction of the visible photoluminescence leads to the observed signal in the NIR detection window. This effect is somewhat diminished by the reduced detector sensitivity and lower grating efficiency for visible photons.

When the microwave irradiation is switched on, the signal significantly changes: the visible emission decreases, while the true NIR emission in the $1042-1200$~nm range increases. This behavior follows from the level scheme illustrated in Fig.~\ref{fig:Fig1_structure_Jablonski_data}a). Microwave excitation redistributes the population from the optically polarized $m_S=0$ state to the $m_S=\pm 1$ states. Since the intersystem crossing preferentially populates the singlet levels from the $m_S=\pm 1$ states, the NIR emission increases, while the visible fluorescence decreases.

The ODMR signal is obtained by taking the difference between the spectra with and without microwave irradiation, as shown in Fig.~\ref{fig:Fig1_structure_Jablonski_data}d. The same signal can also be measured using lock-in detection, which provides a drastically better signal-to-noise ratio.

\begin{figure}[H]
        \centering  
        \includegraphics[width=\linewidth]{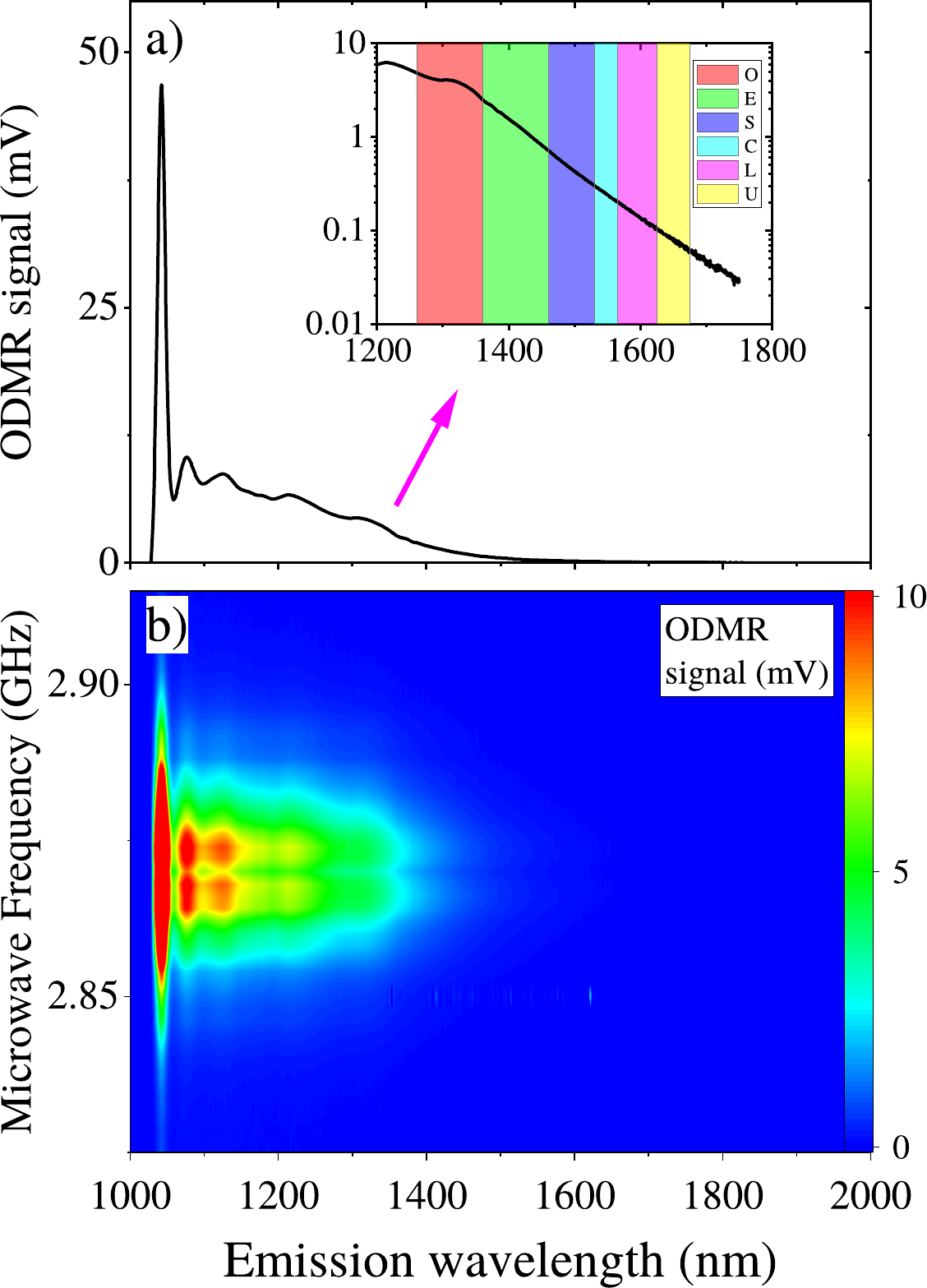}
        \caption{a) Wavelength-resolved ODMR spectra of NV centers measured with a $1000$ nm long-pass filter, showing $1042$ nm zero-phonon peak and some other sidebands. The visible emission is clearly filtered out. The inset shows the signal on a logarithmic scale, demonstrating that it persists up to $1800$ nm, where it reaches the noise floor. The telecommunication wavelength bands are also indicated with colored bars. b) The ODMR map shown as a function of the microwave irradiation frequency. It is constructed from ODMR spectra acquired in individual microwave frequency measurements. Although the maximum of the data is $50$ mV, the displayed maximum is set to $10$ mV to enhance the visibility of weaker signals in the colored contour plot.}
        \label{fig:Fig2_NIR_ODMR}
\end{figure}

Although the use of the $600$~nm long-pass filter clearly demonstrates the opposite sign of the visible and NIR ODMR signals, it limits the observation of the longer-wavelength part of the spectrum. Therefore, in Fig.~\ref{fig:Fig2_NIR_ODMR}a, we show NIR ODMR spectra recorded with a $1000$~nm long-pass filter under continuous microwave irradiation at a fixed $2.874$~GHz frequency. The inset, plotted on a logarithmic scale, shows that the ODMR signal persists up to ${\sim}1800$~nm, thus covering the full spectral range relevant for fiber-optic telecommunications. The corresponding telecommunication bands are also indicated \cite{Agrawal2010FiberOptics}.

From a series of ODMR spectra taken at different microwave frequencies (with a step size of $1$~MHz), one can construct ODMR maps. These maps show that the observed ODMR features are essentially the same across the full wavelength range and for different microwave frequencies, indicating a common physical origin of the signal.

\begin{figure}[H]
        \centering  
        \includegraphics[width=\linewidth]{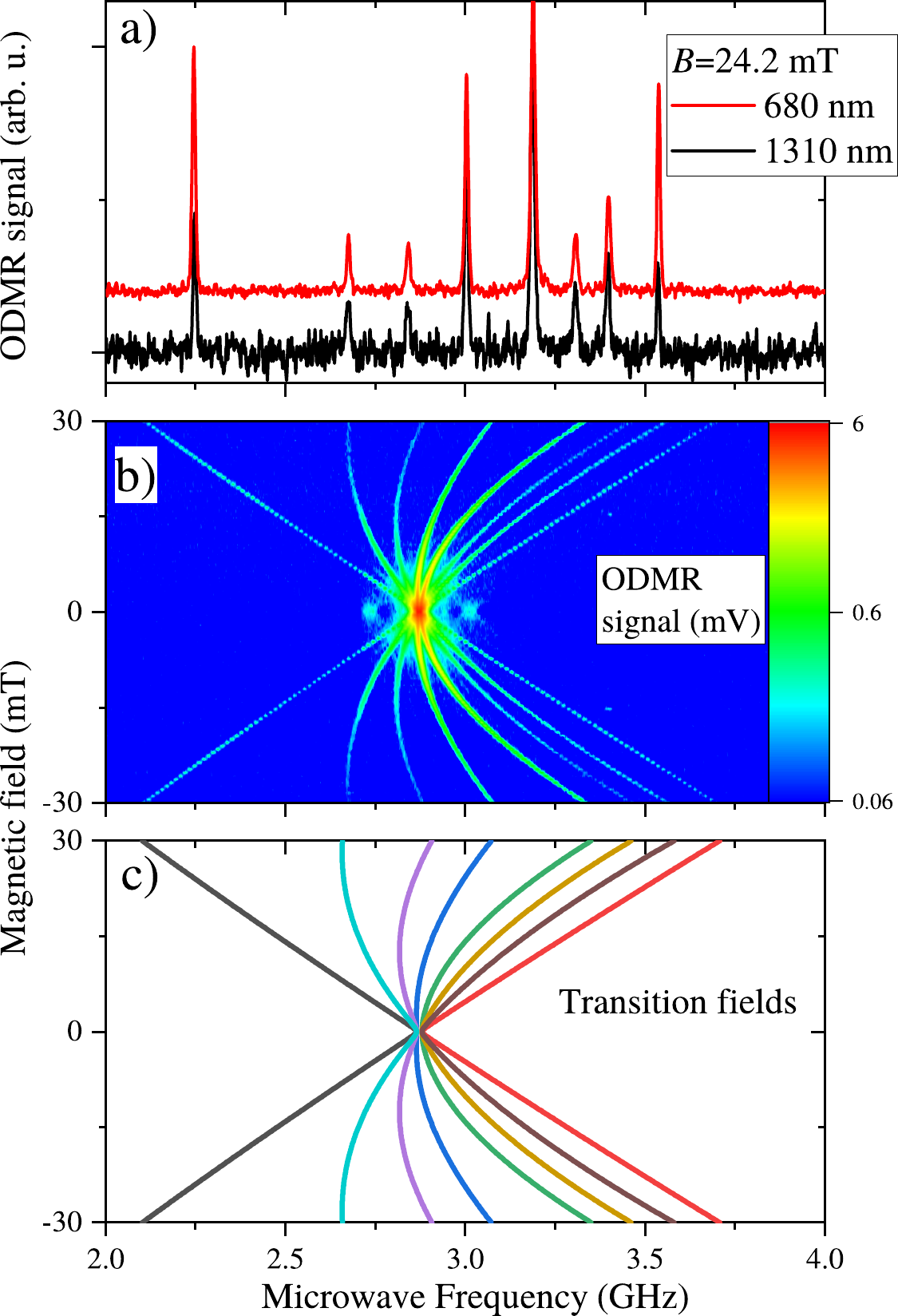}
        \caption{a) ODMR signal of NV centers in a magnetic field of $24.2$ mT as detected in the visible (at $680$ nm) and in the near-infrared (at $1310$ nm). b) Magnetic field-dependent ODMR map as detected in the near-infrared at $1310$ nm (shown on a logarithmic scale for better contrast). The data was obtained by compiling several individual frequency-swept datasets at given magnetic fields. c) calculated magnetic field dependent energy dispersion data for all four possible orientations when the magnetic field points in the $\vartheta=16.9^\circ$, $\varphi=61^\circ$ direction with respect to the NV-axis.}
        \label{fig:Fig3_Magn_field_Dispersion}
\end{figure}

In order to test whether the spin information is preserved in the NIR emission, we compare ODMR spectra detected at different wavelengths. Figure~\ref{fig:Fig3_Magn_field_Dispersion} displays the magnetic-field-dependent ODMR spectra obtained from the visible fluorescence at $680$ nm and from the $1310$~nm singlet emission in a finite magnetic field. In addition, a full \emph{ODMR map} is shown in Fig. ~\ref{fig:Fig3_Magn_field_Dispersion}, which is constructed from individual frequency-swept ODMR spectra in different magnetic fields. 

The two experiments show the same resonance positions and field dependence within the measurement accuracy. Although the signal-to-noise ratio is somewhat inferior at $1310$ nm, the ODMR features remain clearly resolved. This indicates that the same spin transitions are observed in all detection channels.

For comparison, we calculate the expected transition frequencies using the ground-state spin Hamiltonian of the NV center \cite{Doherty2013}:
\begin{equation}
    H = D S_z^2 + E (S_x^2 - S_y^2) + g \muB \mathbf{B} \cdot \mathbf{S},
\end{equation}
with the axial-zero field splitting parameter, $D$, and the transverse zero-field splitting parameter, $E$, which describes the doubling of the ODMR peak in zero magnetic field and is due to strain. The $g=2.0026$ is the $g$-factor. The calculated magnetic-field dependence is shown as solid lines in Fig.~\ref{fig:Fig3_Magn_field_Dispersion} using $D=2872.3~\text{MHz}$, $E=7.44~\text{MHz}$, and a magnetic field forming an angle of $\vartheta=16.9^\circ$, $\varphi=61^\circ$ direction with respect to an NV-axis. The theoretical curves show an outstanding agreement with the experimental data for both detection wavelengths (the ODMR map for the visible data is presented in the Supplementary Information).

These results show that the ODMR signal detected in the NIR follows the same spin transitions as the visible fluorescence. This demonstrates that the spin information is transferred to the infrared emission via the spin-dependent ISC and remains accessible even at wavelengths extending into the telecom range.

The observation of a persistent ODMR contrast up to ${\sim}1800$~nm is relevant in view of fiber-based applications, especially at $1310$ and $1550$ nm, where there is no chromatic dispersion and the optical losses are minimal, respectively \cite{Agrawal2010FiberOptics}. Therefore, the results indicate that NIR detection can serve as an alternative readout channel for NV centers in schemes where compatibility with optical fiber systems is essential.

\section{Conclusions}

We have demonstrated wavelength-resolved ODMR of the infrared singlet emission of NV centers in diamond over a broad spectral range extending up to ${\sim}1800$~nm. The observed ODMR contrast at $1042$~nm and its persistence across the near-infrared range show that spin information is preserved in the singlet emission channel. The spectral behavior and magnetic-field dependence of the signal are consistent with the established spin-dependent ISC mechanism, indicating that the infrared emission carries the same spin contrast as the visible fluorescence, albeit with opposite sign.

The wavelength-resolved measurements further show that the ODMR features are essentially identical across the investigated spectral range and for different microwave frequencies, confirming a common physical origin of the signal. These results establish infrared ODMR as a viable alternative readout scheme for NV centers.

Our observation of ODMR contrast persisting into the U-band (up to $1800$ nm) marks a shift in how the NV singlet manifold is perceived -- from a \emph{dark} parasitic loss channel to a direct, conversion-free telecommunication port, making it directly applicable. This infrastructure-compatible readout scheme streamlines the integration of diamond-based quantum nodes into existing fiber-optic and CMOS architectures by eliminating the overhead of external nonlinear optics.

\section{Acknowledgment}
Work supported by the National Research, Development and Innovation Office of Hungary (NKFIH), and by the Ministry of Culture and Innovation Grants Nr. 149457, 2022-2.1.1-NL-2022-00004. The Cooperative Doctoral Programme (KDP-2025) is also acknowledged for support. Project No. 2025-2.1.2-EK\"OP-KDP-2025-00005 has been implemented with the support provided by the Ministry of Culture and Innovation of Hungary from the National Research, Development and Innovation Fund, financed under the EK\"OP\_KDP-25-1-BME-8 funding scheme.

\section{Methods}

High-pressure--high-temperature (HPHT, Type Ib) grown single crystal diamonds from Element Six were used for the measurements. The samples were subjected to neutron irradiation in a nuclear reactor in accordance with our previous work \cite{KollaricsCarbon}, followed by annealing at $1000~^\circ\text{C}$ for two hours. The resulting NV center concentration of about ${\sim}1$ ppm was determined by electron paramagnetic resonance (EPR) spin-counting technique against a copper sulfate pentahydrate standard (CuSO$_4\cdot5$\,H$_2$O) using a commercial Bruker Elexsys E500 continuous wave electron spin resonance spectrometer.

\begin{figure}[H]
        \centering
        \includegraphics[width=\linewidth]{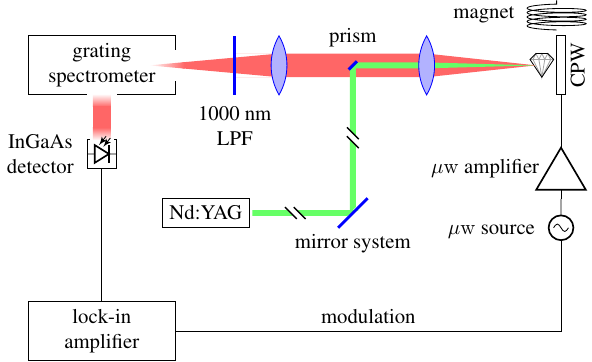}
        \caption{Schematics of the wavelength-resolved ODMR instrument. A $532$ nm laser excites the diamond sample through a focusing lens. The same lens collects the photoluminescent photons, which are focused on the input slit of a Czerny-Turner spectrograph with another lens, followed by a long-pass filter (we used $532$, $600$, or $1000$ nm filters). The spectrograph is equipped with a single-channel PMT for visible and an InGaAs detector for NIR detection. The sample is inside an electromagnet on a conductor-backed CPW, which acts as a microwave antenna. }
        \label{fig:Fig4_block_diagram}
\end{figure}

The block diagram of the home-built near-infrared (NIR) optically detected magnetic resonance (ODMR) spectrometer is shown in Fig.~\ref{fig:Fig4_block_diagram}. The instrumentation is detailed in previous articles \cite{Negyedi2017, Palotas2020} and is briefly summarized herein. The sample was illuminated using a frequency-doubled Nd:YAG laser operating at $532$ nm with a maximum power of $200$ mW. The emitted light was collected and focused onto the spectrograph by an achromatic doublet pair (Thorlabs MAP1030100-B). The spectrograph (Horiba Jobin Yvon iHR320) was equipped with a photomultiplier tube (Hamamatsu R2658P) and an InGaAs photodiode detector (Horiba DSS-IGA010L), covering the visible (VIS) and near-infrared (NIR) spectral ranges. The detector signal is first fed to a transimpedance amplifier and then passed to a lock-in amplifier (Stanford Research Systems SR830) for phase-sensitive detection. For the photoluminescence spectra, an optical chopper (Stanford Research Systems SR540), with typical chopping frequencies below $200$ Hz, was utilized. This is necessary due to the relatively long spin-lattice relaxation time of the NV system.

The sample was mounted on a conductor-backed coplanar waveguide (CPW), which acts as a broadband microwave antenna. To record the ODMR signal directly, the output of the microwave source (HP83751A) was amplitude-modulated by the TTL signal of the lock-in amplifier at a given local oscillator frequency, and the optical signal was synchronously demodulated at this frequency. The microwave signal was optionally amplified to $40$ dBm ($10$ W) using a power amplifier (Kuhne Electronic GmbH, KuPA 270330-10A, Gain $40$ dB), and the signal was properly terminated to $50~\Omega$ to avoid reflections. An external magnetic field was applied using a JEOL electromagnet and a LakeShore DC power supply, with the field strength monitored by an AC Hall probe.

The unique feature of our instrument, compared to conventional ODMR spectrometers, is the ability to resolve the wavelength of the detected light and its sensitivity in the NIR range up to $2000$ nm. We use a set of long-pass filters (LPF) at the input of the monochromator according to the desired detection range: $532$ nm, $600$ nm, $800$ nm, or $1000$ nm (Semrock RazorEdge $532$ nm, Thorlabs FEL0600, 800, and 1000). The detection of a grating-type monochromator is inevitably degenerate for second-order diffraction wavelengths; thus, when using the $600$ nm LPF, the otherwise significant visible photoluminescence (PL) signal of the NV centers also appears above $1200$ nm in the NIR measurements, i.e., second-order diffraction of shorter wavelengths overlaps with the first-order NIR signal.

\clearpage
\newpage
\onecolumngrid

\section{Supporting Information: \\Telecom-Wavelength-Compatible Quantum Information Transcription Using Nitrogen-Vacancy Centers}

\renewcommand{\appendixname}{S}
\renewcommand{\thesection}{S}
\renewcommand\thefigure{\thesection\arabic{figure}}
\setcounter{figure}{0}
\renewcommand\thetable{\thesection\arabic{table}}
\setcounter{table}{0}

\appendix

\begin{figure}[hp]
        \centering  
        \includegraphics[width=\linewidth]{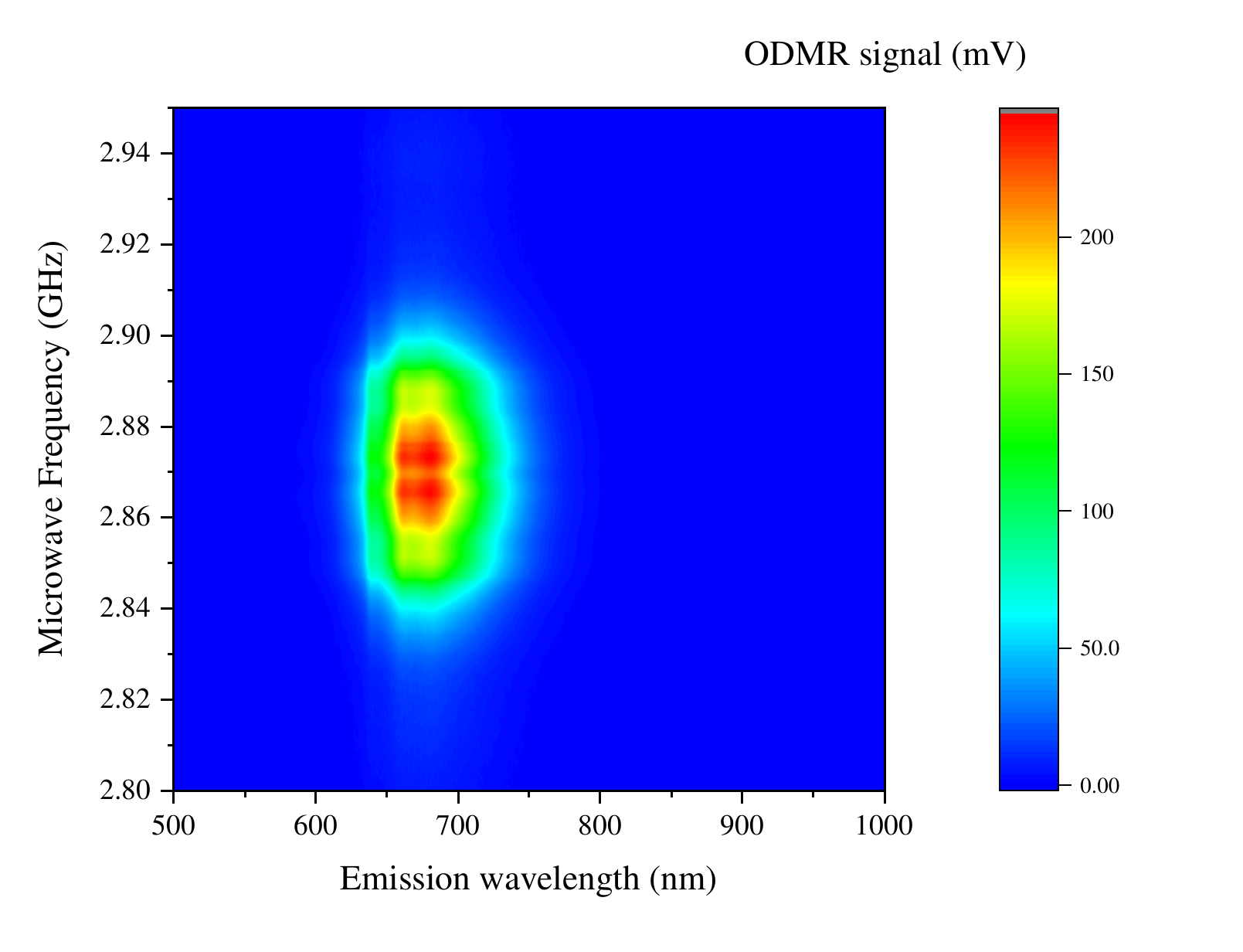}
        \caption{Wavelength resolved ODMR map of NV centers in the visible regime. A $532$ nm long-pass filter was used to obtain the data.}
        \label{fig:Fig1_SM_ODMR_map.pdf}
\end{figure}

Fig. \ref{fig:Fig1_SM_ODMR_map.pdf} shows the ODMR map in the visible region. A $532$ nm long-pass filter was used to obtain this data, and the map consists of several individual ODMR spectra acquired with $1$ MHz microwave frequency steps. We note that the data in Fig. \ref{fig:Fig1_SM_ODMR_map.pdf} is inverted as the ODMR signal is negative in the visible range, i.e., in the presence of microwaves, the PL intensity decreases. 

\begin{figure}[htp]
        \centering
        \includegraphics[width=\linewidth]{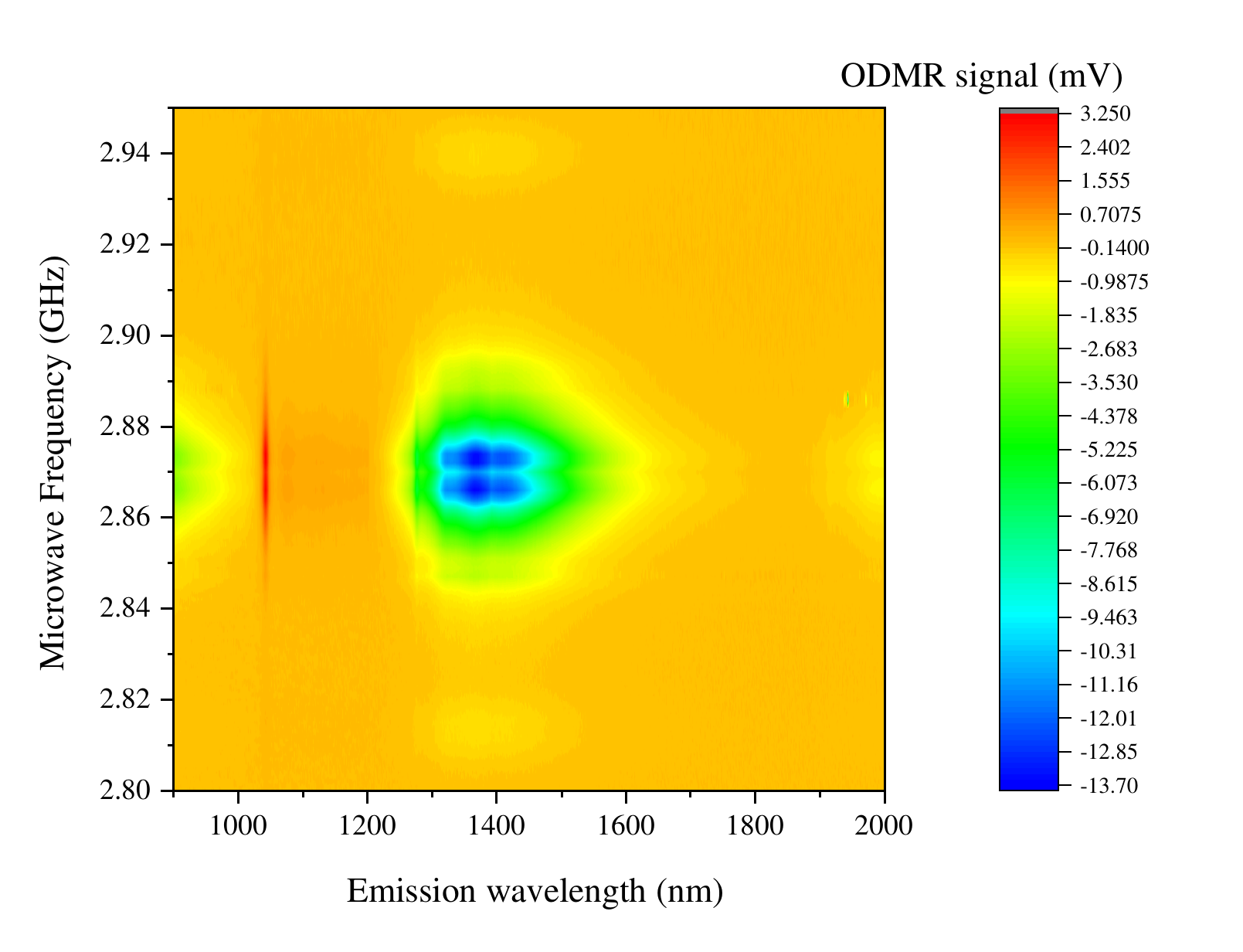}
        \caption{Simultaneously detected near-infrared and visible ODMR signals using a NIR optimized detector and a $600$ nm long-pass filter. The $1042$ nm singlet-singlet transition as well as the visible ODMR signal are both visible with an opposite sign (the latter is negative). The visible signal is observed due to the $2^{\text{nd}}$ harmonic of the grating.}
        \label{fig:Fig2_SM_NIR_ODMR_600nm_filter.pdf}
\end{figure}

Simultaneous detection of the near-infrared (NIR) and visible optically detected magnetic resonance (ODMR) signals was performed using a detection scheme optimized for infrared sensitivity. A NIR-optimized detector in combination with a $600$ nm long-pass filter enables access to the emission associated with the singlet–singlet transition near $1042$ nm, while still transmitting a residual component of the visible fluorescence. As shown in Fig. \ref{fig:Fig2_SM_NIR_ODMR_600nm_filter.pdf}, both the infrared and visible ODMR signals are observed concurrently. Notably, the two signals exhibit opposite contrast: the ODMR signal in the NIR channel appears with positive contrast, whereas the visible contribution is negative, consistent with the conventional fluorescence-based ODMR response of the NV center. The presence of the visible signal in the nominally infrared detection channel is attributed to the second-order diffraction of the grating in the spectrometer, which maps shorter-wavelength visible photons into the same detection window as the NIR emission. This artifact, while typically undesirable, provides an internal reference (as discussed in the main text) that confirms the simultaneous acquisition of both optical channels and highlights the distinct contrast mechanisms associated with triplet and singlet transitions.

\begin{figure}[htp]
        \centering
        \includegraphics[width=\linewidth]{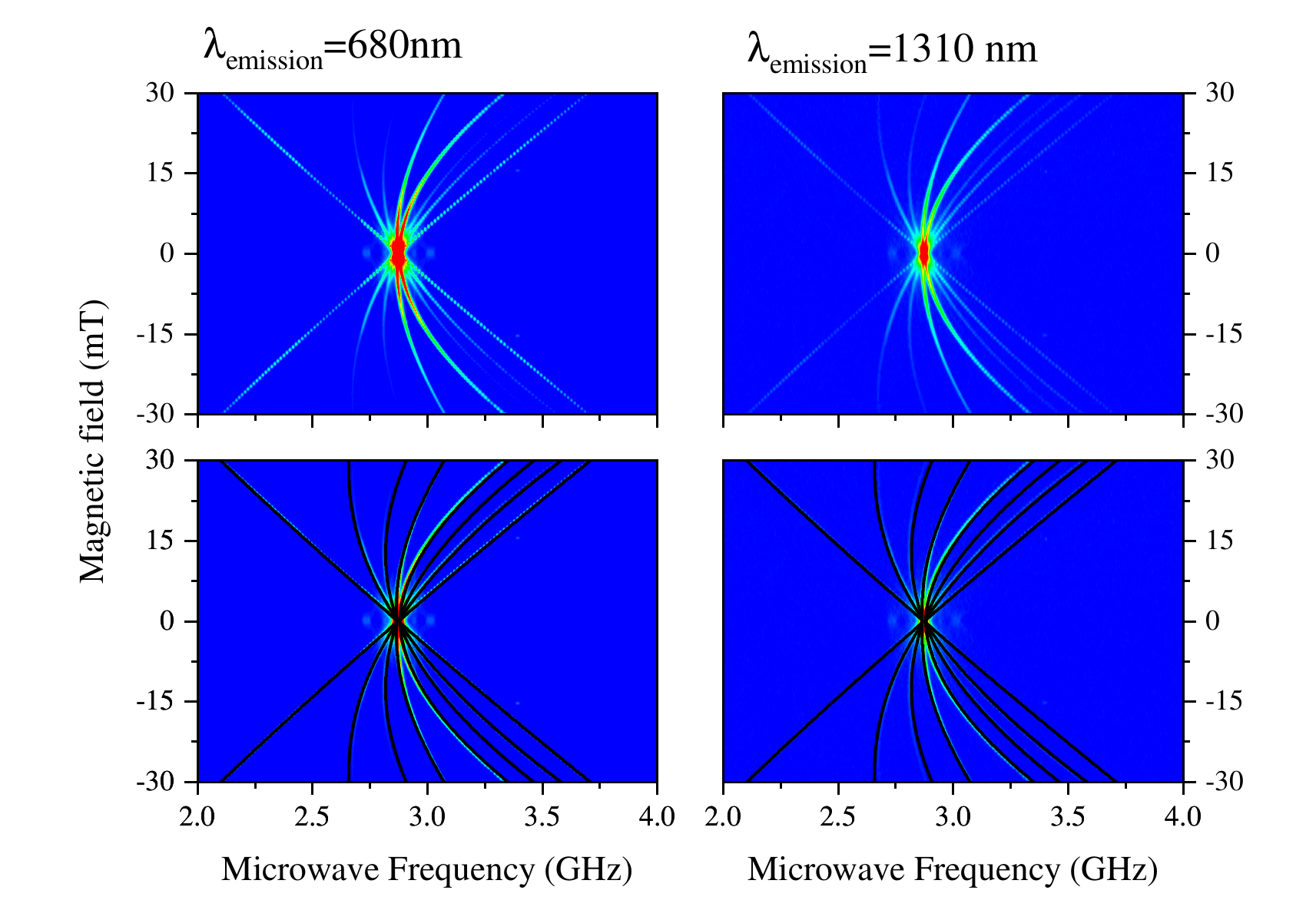}
        \caption{ODMR maps recorded at two emission wavelengths, $\lambda_{\mathrm{em}} = 680~\mathrm{nm}$ (left) and $\lambda_{\mathrm{em}} = 1310~\mathrm{nm}$ (right). The top panels show the raw ODMR intensity as contour plots as a function of microwave frequency and magnetic field. The bottom panels display the same data overlaid with the calculated eigenvalues obtained from diagonalization of the spin Hamiltonian, demonstrating excellent agreement between experiment and model.}
        \label{fig:Fig3_SM_VIS_vs_NIR_ODMR}
\end{figure}

ODMR measurements were performed at two emission wavelengths, $\lambda_{\mathrm{em}} = 680~\mathrm{nm}$ and $\lambda_{\mathrm{em}} = 1310~\mathrm{nm}$. The resulting ODMR intensity maps, recorded as a function of microwave frequency and magnetic field, reveal characteristic resonance branches in both spectral ranges as shown in Fig. \ref{fig:Fig3_SM_VIS_vs_NIR_ODMR}. While the raw data exhibit similar overall structures, a quantitative comparison requires modeling of the underlying spin Hamiltonian. To this end, the Hamiltonian was diagonalized, and the corresponding eigenvalues were overlaid on the experimental maps, showing excellent agreement across the full field and frequency range.

The visible emission at $680~\mathrm{nm}$ was detected using a $532~\mathrm{nm}$ long-pass filter. In contrast, the $1310~\mathrm{nm}$ measurements were carried out using a near-infrared (NIR) optimized spectrometer and detector system, equipped with a $1000~\mathrm{nm}$ long-pass filter. The consistency of the ODMR features observed in the visible and NIR regimes confirms that both detection channels probe the same underlying spin transitions.

\end{document}